# Enhancing Information Systems Security in Educational Organizations in KSA through proposing security model

### Hussain A.H. Awad and Fadi M. Battah

**Department of Computer and Information Technology, King Abdulaziz University, Faculty of Science and Arts- Khulais Jeddah, KSA**

**Department of Computer and Information Technology, King Abdulaziz University, Faculty of Science and Arts- Khulais Jeddah, KSA**

### Abstract
It is well known that technology utilization is not restricted for one sector than the other anymore, Educational organizations share many parts of their information systems with commercial organizations.
In this paper we will try to identify the main characteristics of information systems in educational organizations, then we will propose a model of two parts to enhance the information systems security, the first part of the model will handle the policy and laws of the information system, the second part will provide a technical approach on how to audit and subsequently maintain the security of information system.

## 1. Introduction

According to Encyclopedia Britannica; a university is an institution of higher education and research, which grants academic degrees in a variety of subjects. A university is a corporation that provides both undergraduate education and postgraduate education. The word university is derived from the Latin universitas magistrorum et scholarium, roughly meaning "community of teachers and scholars."

With the higher competitive environment around the world, educational organizations are not saving any effort to provide the best educational experience. This effort includes employment of latest technologies available; from the entrance of computers mid 20[th] century up to the outsourcing of complex Enterprise Resource Planning systems and usage of cloud computing. This usage along side with the expanded branches of educational organizations have presented new challenges, the virtual private networks, wide area networks and usage of web interfaces all together made the educational organizations target as same as any other organization on the cyber space.

According to WhiteHat website security statistics report 2011 *''Most websites were exposed to at least one serious vulnerability every day of 2010, or nearly so (9–12 months of the year). Only 16% of websites were vulnerable less than 30 days of the year overall.''* And *''71% of Education, 58% of Social Networking, and 51% of Retail websites were exposed to a serious vulnerability every day of 2010''.*
In this paper we will tackle the issue of Information Systems safety in educational organization, considering King Abdulaziz University as a case study and propose a two tier model for enhancing the security of information system in educational organizations..

## 2. Information Systems Security

''Information system security relates to the adequacy of management controls to prevent, avoid, detect and recover from whole range of threats that could cause damage or disruption to computer systems.'' (Pattinson, 2008), the process of information security cannot provide a complete prevention, avoidance, detection and recovery from the threats over it (Singh, 2008). And any self aware Information Systems Management realize that; but the fact that any action of information security management can help to reduce these factors gives that motive to embrace all strategies, models and techniques to achieve that.
We can identify the main process of the information system security in the following diagram based on the previous definition:

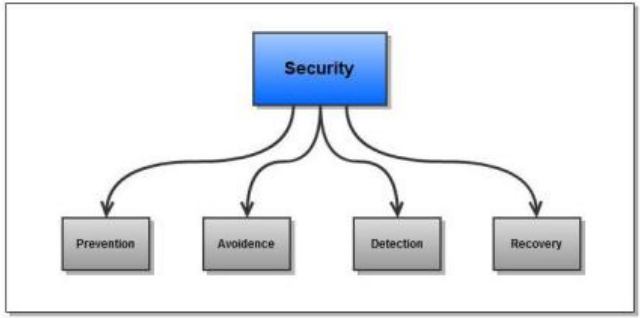

Figure 1 The Main Process of IS Security

## 2. Universities and Information Systems

"A university information system has to provide information about research and scientific cooperation offers, education and further education capabilities."(Kudrass, 2006). Information systems in universities can be considered more complex than the usual information systems used in commercial organization. But still it must pay the same attention to its customers (students and members) (Luo and Warkentin, 2004).

2.1 King Abdulaziz University System

The complexity In the King Abdulaziz University information system is relative to The Land Grant University System (Chae and Poole, 2009) and comprised of the following main components:

A- The Students Systems that include
    1- On Demand University Services (ODUS)
    2- Electronic Report System (ERS)
    3- Virtual Classes System (CENTRA)
    4- Electronic Management of Education System (EMES)
B- The Academic Systems that include
    1- On Demand University Services (ODUS)
    2- Academic Affairs System (SMART)
    3- Academic Services for Higher Education
    4- Anjez system for human resources, financial management and memo's.
    5- Performance Management System (PMS)
    6- Evaluation System
C- Management Systems that include:
    1- Anjez system
    2- Employment system
    3- Decisions and memos system.

    4- Performance Management System (PMS)

The above systems can be viewed by each member (Student, Academic or Employee) depending on his unique number and password. Those information systems are supported by the infrastructure used in the university to provide connectivity amongst campuses in the kingdom and provide internet services for the users.

2.2 Risk Analysis

The main subject of this analysis is to identify to what extent information is subject to change or exposure in this system (Pattinson, 2008), this identification poses the questions of what the possibility of such threats? And what are the expected losses upon such change or exposure?

In real situation it hard to answer these questions, according to (Janczewski, 2009) the reasons behind that are:

- The cost and duration to collect such probabilities may be so huge that job will not be acceptable to management.

- Attacks never happened, but they may happen in the future, so there is no reliable loss of information.

Another factor that raises the risk is the nature of educational organizations, since university standings could suffer greatly from any sort of data manipulation or exposure.

We can provide theoretical risk assessment depending on the nature of the system; the system in the university depends mainly on unique user names and password and on data acquisition upon transfer, especially among long distance branches.

Further investigation to risk analysis requires elaborate software applications which are not our purpose for this paper.

## 3. Information Systems Security Model

In this section we will tackle the main issue of our paper. Although methodologies are yet till now still not completely mature (Torres, 2009) and some models were proposed to for enterprise security systems (Mazumdar, 2009); we will propose a two part model that would help in the enhancement of information security in universities. This model is divided into two parts; this first is the Policy part, and the second is the Auditing part.

## 3.1 Policy

We will use the policy based management (Perez, 2006) in order to create an Information System Security (ISS) policy will be aimed to provide the supporting background to regulate the usage by members of the system in a way that raises the system safety and information rigidity to aim in the whole purpose of accomplishing the university strategic goals.

The ISS Policy will be made of the following main points:

1- Statement of Purpose: the statement will in coherence with flexibility, political simplicity and criterion orientation (Baskerville and Siponen, 2002), (Ghormley, 2009).

2- Policy Application Plan: since we are applying new policy to an existing system; a plan must be made for the process of applying this policy, this plan must be governed by time schedule. And it must be made clear that all members comply with this policy (Hinson, 2009).

3- Policy and Standards

1- Overview

2-Responsibility Delegation: Identify the responsibilities for all members by providing security agreement upon using the organizations system. And identify the main manager of the security by appointing Chief Security Officer (CSO) and enable him to form his department accordingly.

3-Contingency Policy: provide plans and walkthrough in case of information disaster, this include performing risk analysis and assessment, business impact analysis, prepare and apply backup and recovery strategy and maintain information update in the policy.

4- Copyrights policy: all members must be informed and agree on the copyright policy for the resources provided in the educational organization system, these copyrights abide to local and international laws. Also they are in line with information technology politics (Petrides, Khanuja-Dhall, & Reguerin, 2006)

5- Help Disk: provide a help disk and hotline to help members in any case of data loss either physical or digital, malfunction of tools, malware and viruses and cases of physical robbery to equipment that holds sensitive data.

6- Accounts and Passwords: members must agree to protect their accounts and not to share any personal information with other members or outside individuals; also the organization must enforce high security policy for selecting personal account passwords, such as using special characters and password change periods.

7- Equipment and Facilities Security: the organization must provide all required equipment, personnel and facilities to protect the main hardware of the system. Also any part of the system design can be count as equipment (Janczewski, 2009).

8- Networking: the organization must provide all necessary tools and equipment to deliver communication network to all university facilities, in the case of KAU this includes colleges outside the city; which means using the latest technology in fiber optics and wireless communication in order to keep all members connected through the main network.

9- Web Security: the organization must work on employing the best security for the webpage since it is the interface used outside the university network and used to access mostly all data.

10- Personal Computers and Laptops: the organization must govern the connectivity to the network using several technologies such as active directory and router identification of computer by MAC address, this will help keep the network secure in case of any attempt to connect unauthorized computer to it.

11- Data Backup: the organization well provides a scheduled process for backup, alongside with providing data containers for the members to store their information on the university equipment as redundant copy.

12- Inventory and turn over policy: the department must write down policy for keeping track of all equipment and the method of upgrades, maintenance and replacement in case of damage, this includes advanced methods of data disposal of old data storage.

13- Users Ethics: a policy of ethical usage of the resources and data provided through the university network must be prepared and signed by all members of the organization. This will help in raise the value of IS in general (Kizza, 2008).

This policy can be furthered in relation with ISO 2700 (Calder, 2006) for information security management system and must be checked regularly in order to make sure it is up-to-date with the latest standards and technologies (Tong, 2009)

## 3.1 Auditing

This is the second part of our model, this part is concerned of the actual daily vulnerabilities that could happen, in order to do that an audit of our system must be done periodically, the less the period the better; but due to the nature of our organizations it is hard to conduct such tests and audits in a way that could compromise the system.

In this part we propose a schema for the auditing process, together with the policy this will provide a solid infrastructure for information system security in educational organizations, this process is aimed for the internal use of the information systems department under the approval of higher management only.

### 3.1.1    The Plan

Any process needs planning, and so in our auditing phase, where this plan will help in the total overview of our process and evaluation of it at the end of each audit process.

A: Establish Goals.

We must at first determine the goals of this process, mainly is to find and secure any vulnerability in our system, these could happen due to: human, software or hardware factors, all must be considered. This plan also must identify clearly the time schedule of this process in order to maintain daily processes ongoing and not interrupting them.

B: Identify Targeted System.

Information about the targeted system is very important in order to identify the size of threats and vulnerabilities, networking protocols, networking schemes, operating systems, management system used and even the forum management software all must be identified.

C: Create Audit Standards. International standards and regulations could be taken in consideration (D'Arcy and Hovav, 2009).

D: Select Security Assessment Tools.

### 3.1.2    Methodology

A: Preparation.

This is the data collection stage, in which the auditing team works on any information available about the organization.

B: Scanning the system.

Starting with the default ports used then expanding to least used ports; this will help on identifying the entrance to the system.

C: Classify vulnerabilities

D: Break in the system.

Using all information gathered; the team will enforce penetration to the system in order to start the next phase of acquiring information.

### 3.1.3    Using Social Engineering to Acquire Data

Other than the machines and software; humans are main component in any system (Kuusisto, 2009), and such compenet need some research (Ada, 2009).After using information and wholes of the system; the next step is acquiring data from its containers, this needs more effort using social engineering in order to override: Physical security, Passwords, Network Security, Data Base and Storage Systems

### 3.1.4    Curtain Down

In this stage all information about the system is analyzed, all holes and vulnerabilities are acknowledged in order to create a plan to plug all the security holes of the system.

## 4. Summary

In this paper we have proposed a model that can be used to enhance information system security in educational organizations, this model is divided into two main parts, the first part is the policy making and publishing; this part is done thoroughly for the first time then a review and enhancement is done. The second part is the security auditing process; this part must be performed periodically keeping in mind the updates in the software used and any new parts to the system.

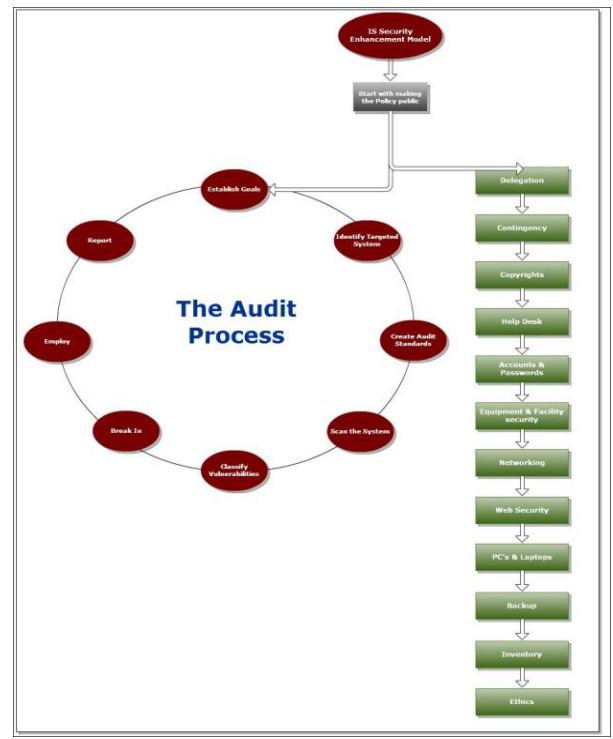

Figure 2 Information Security Enhancement in Educational Organizations Model

**Hussain A.H Awad** received his B.Sc. in Business Administration from Mutah University in Jordan, and obtained his M.Sc in Management Information Systems from Amman Arab University in Jordan, and Ph.D. in Management Information Systems from the Arab Academy for Banking and Financial Sciences in Jordan. He is now an Assistant Professor at the Department of Management Information Systems in KING ABDULAZIZ UNIVERSITY, Faculty of Science and Arts - Khulais, Jeddah - Kingdom of Saudi Arabia. His current research interests are supply chain management, IS security, and information retrieval.

**Fadi M. Battah** Holds a BSc. In Computer Science and M.Sc. in Information Technology Management from University of Sunderland, UK. And has worked as an IT Manager in private sector and now working as a Lecturer at the Department of Management Information Systems in KING ABDULAZIZ UNIVERSITY, Faculty of Science and Arts - Khulais, Jeddah - Kingdom of Saudi Arabia. His current research interests are Information Security, Networking Management, Outsourcing and Information Technology Management.